\documentclass[aps,prb,amsmath,amssymb,twocolumn,superscriptaddress]{revtex4-2}

\setcitestyle{numbers,square}

\usepackage{graphicx}
\usepackage{dcolumn}
\usepackage{bm}
\usepackage{hyperref}

\begin{document}

\title{Light-induced ultrafast dynamics of spin crossovers in LaCoO$_3$}

\author{Yu.S. Orlov}
\affiliation{Kirensky Institute of Physics, Federal Research Center KSC SB RAS, 660036 Krasnoyarsk, Russia}
\affiliation{Siberian Federal University, 660041 Krasnoyarsk, Russia}
\email{jso.krasn@mail.ru}

\author{S.V. Nikolaev}
\affiliation{Kirensky Institute of Physics, Federal Research Center KSC SB RAS, 660036 Krasnoyarsk, Russia}
\affiliation{Siberian Federal University, 660041 Krasnoyarsk, Russia}

\author{S.G. Ovchinnikov}
\affiliation{Kirensky Institute of Physics, Federal Research Center KSC SB RAS, 660036 Krasnoyarsk, Russia}
\affiliation{Siberian Federal University, 660041 Krasnoyarsk, Russia}

\date{\today}

\begin{abstract}
Ultrafast quantum dynamics relaxation of a photoexcited state in a strongly correlated spin crossover system LaCoO$_3$ under a sudden perturbation is considered with the density matrix generalized master equation. The magnetization and cobalt-oxygen bond length oscillations were found. The evolution of the electronic band structure during relaxation is calculated in the framework of the LDA+GTB method.
\end{abstract}

\maketitle


In LnCoO$_3$ the ground state of the Co$^{3+}$ ion is known to be in the low spin (LS) ${}^1A_{1g}$ state with $S=0$, so the ground state is non magnetic. It is known also that the lowest excited high spin (HS) ${}^5T_{2g}$ term is separated from the LS one with rather small excitation energy (a spin gap), which is minimal for La (about 10~meV) and increasing for more heavy Ln ions \cite{1}. That is why for LaCoO$_3$ spin state transition is known for a long time with heating \cite{2,3,4}. This transition results from the thermal population of the HS- terms. Recently a new direction in the spin crossover study appears due to new experimental possibilities to switch a LS- state into the HS one by femtosecond irradiation and then to study the spin crossover dynamics by the $X$-ray spectroscopy methods with time resolution. Using $X$-ray free electron laser \cite{5,6} such dynamics has been studied in metal-organic complexes Fe(phen)2(NCS)2. For LaCoO$_3$ a first time-dependent LS-HS dynamics has been discussed theoretically in the paper \cite{7} and studied with femtosecond soft $X$-ray spectroscopy in the paper \cite{8} where an ultrafast metallization has been detected.

It is known that the ionic radii of the HS- and LS- states have quite large (about 10\%) difference. It means that excitation from LS- to HS- state results in a remarkable local distortion. Therefore, a multiplicity fluctuation results in a strong electron-phonon anharmonicity. Previously we have studied dynamics in a model with HS/LS ionic states under high pressure that may induce spin crossover \cite{9}. Femtosecond switching from HS to LS was considered in a sudden approximation, and the relaxation of the HS concentration, metal -- ligand bond length and the local magnetization was obtained by a numerical solution of the generalized Master equations. Taking in mind possible ultrafast dynamics experiment we report here the results of the theoretical dynamics of LaCoO$_3$, where LS initial state is excited by a femtosecond pump. With heating LaCoO$_3$ undergoes a smooth insulator -- metal transition, when concentration of the HS- state is above some critical value $n_{HS} = 0.85$, the semiconductor gap is closing \cite{10}. Here we will discuss the LS/HS excitation dynamics, keeping in mind two features of LaCoO$_3$ that can be measured by the $X$-ray spectroscopy methods with time resolution. That is the time dependence of the Co-O bond length, and time dependent metallization.


\begin{figure}
\begin{center}
\includegraphics[width=1.0\columnwidth]{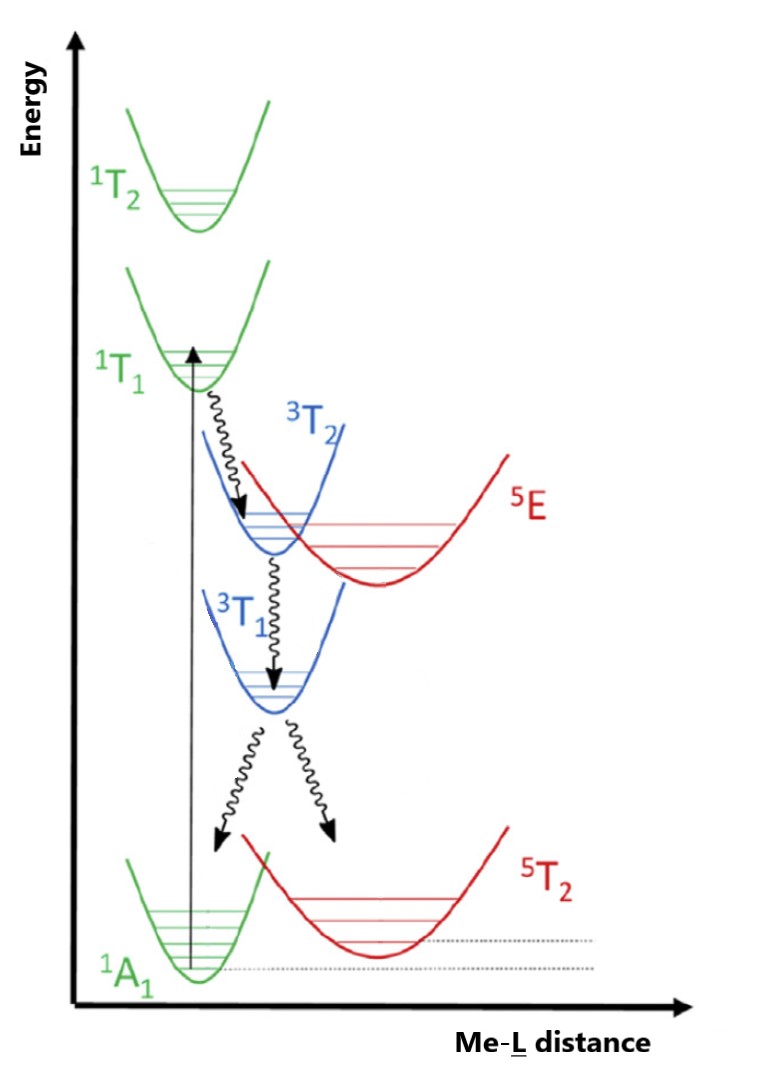}
\caption{\label{fig:1}Sudden photoexcitation from the ground ${}^1A_{1g}$ state to the ${}^1T_{1g}$ state and relaxation of a multielectron system of $d^6$- ions in the ligands crystal field.}
\end{center}
\end{figure}

Using the generalized master equation for the density matrix, we consider the ultrafast quantum relaxation dynamics of a photoexcited state in systems with a spin crossover under a sudden perturbation. The calculations are carried out taking into account the Coulomb and spin-orbit interactions in the framework of the Tanabe-Sugano theory (full multiplet theory) and the electronic-vibrational interaction beyond the adiabatic approximation.

The Fig.~\ref{fig:1} shows photoexcitation from the ground ${}^1A_{1g}$ state to the ${}^1T_{1g}$ state and relaxation process for the $d^6$ electronic configuration of a transition metal ion in an octahedral ligands field. Various channels of relaxation and the appearance of relatively long-lived metastable states are possible.

When the system passes from the light-excited Franck-Condon state to the ground state, the temporal dynamics of the lattice, magnetic and electronic band structure is calculated. At each time step, a self-consistent calculation of the magnetization and occupation numbers of multielectron terms is performed. The electronic band structure is calculated using the LDA+GTB method \cite{11}, where the key role is played by the occupation numbers of multielectron states. For example, for $d^6$- ions, the population of the HS- state of more than 85\% leads to a semimetallic type of band structure, while the ground unexcited LS- state is a nonmagnetic dielectric. The relaxation of the system is usually accompanied by magnetization oscillations and a change in the type of magnetic ordering (photoinduced magnetic transition), since the sign and magnitude of the interatomic exchange interaction in excited states differs from those in the ground state \cite{12}. Due to the strong relationship between the lattice, magnetic, and electronic degrees of freedom in most transition metal oxides, an external action (pumping) on one of the subsystems leads to significant changes and a response in the other. The details of calculations may be found in \cite{9,13}.

\begin{figure*}[h]
\begin{center}
\includegraphics[width=2.0\columnwidth]{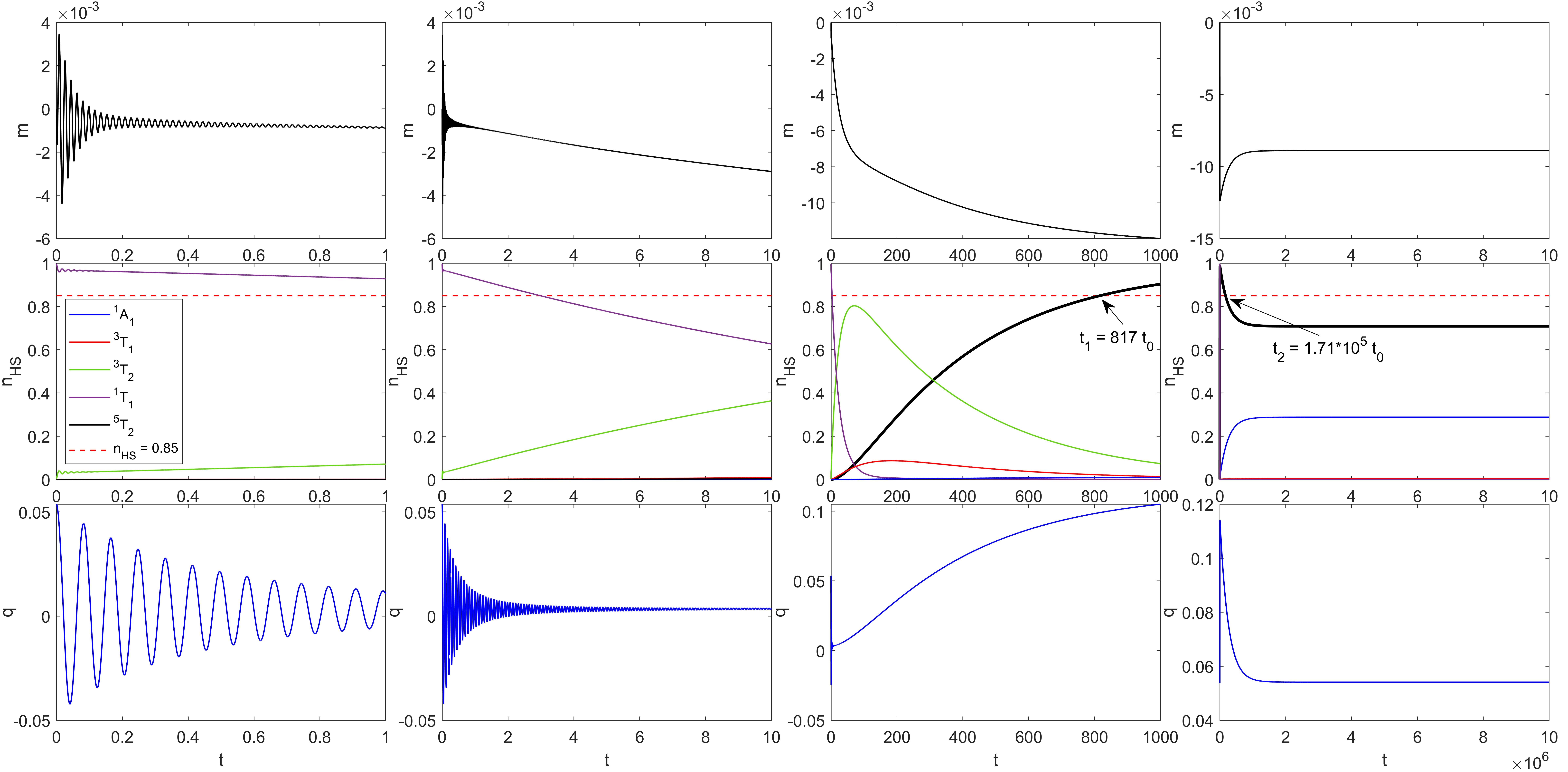}
\caption{\label{fig:2}Quantum relaxation dynamics of the Frank-Condon $^1T_{1g}$ state, photoexcited from the ground LS- state. The upper row shows the magnetization $m = \left\langle {\hat S}^z \right\rangle $ for the sublattice, the middle row shows the term occupancy $n_{HS}$, and the bottom row shows the average value of the normal coordinate operator corresponding to the breathing vibrational mode of ligand's $q$~[\AA]. The red dashed line marks the term occupancy of 85\%. Everywhere, the time along the abscissa is given in units of $t_0 = 10^{-12}$~sec. Calculations were performed for the following parameter values: Racah parameters $B = 1065$~cm$^{-1}$ (0.132~eV), $C = \gamma B$, $\gamma  = 4.808$; spin-orbit coupling constant $\xi  = 400$~cm$^{-1}$ (0.0496~eV); electron-vibration coupling constants linear $g_1 = 2.2$~eV/\AA~and quadratic ${g_2} = 0.5$~eV/\AA$^2$ in the lattice displacement; interatomic exchange interaction $J_{HS - HS} = 2.43$~meV (AFM), $J_{IS - IS} = 1.38$~meV (FM), $J_{HS - IS} = 0.6$~meV (AFM).
}
\end{center}
\end{figure*}

Figure~\ref{fig:2} presents the results of calculation of the quantum relaxation dynamics for $d^6$ electron configuration from the photoexcited Franck-Condon ${}^1T_{1g}$ state at temperature $T = 280$~K. The upper row shows the magnetization $m = \left\langle {{{\hat S}^z}} \right\rangle$ for the sublattice, the middle row shows the term occupancy $n_{HS}$ (the red dashed line marks the occupancy of 85\%), and the bottom row shows the average value of the normal coordinate operator corresponding to the breathing vibrational mode of ligand's $q$~[\AA].

\begin{figure*}[h]
\begin{minipage}[h]{0.49\linewidth}
\center{\includegraphics[width=1.0\linewidth]{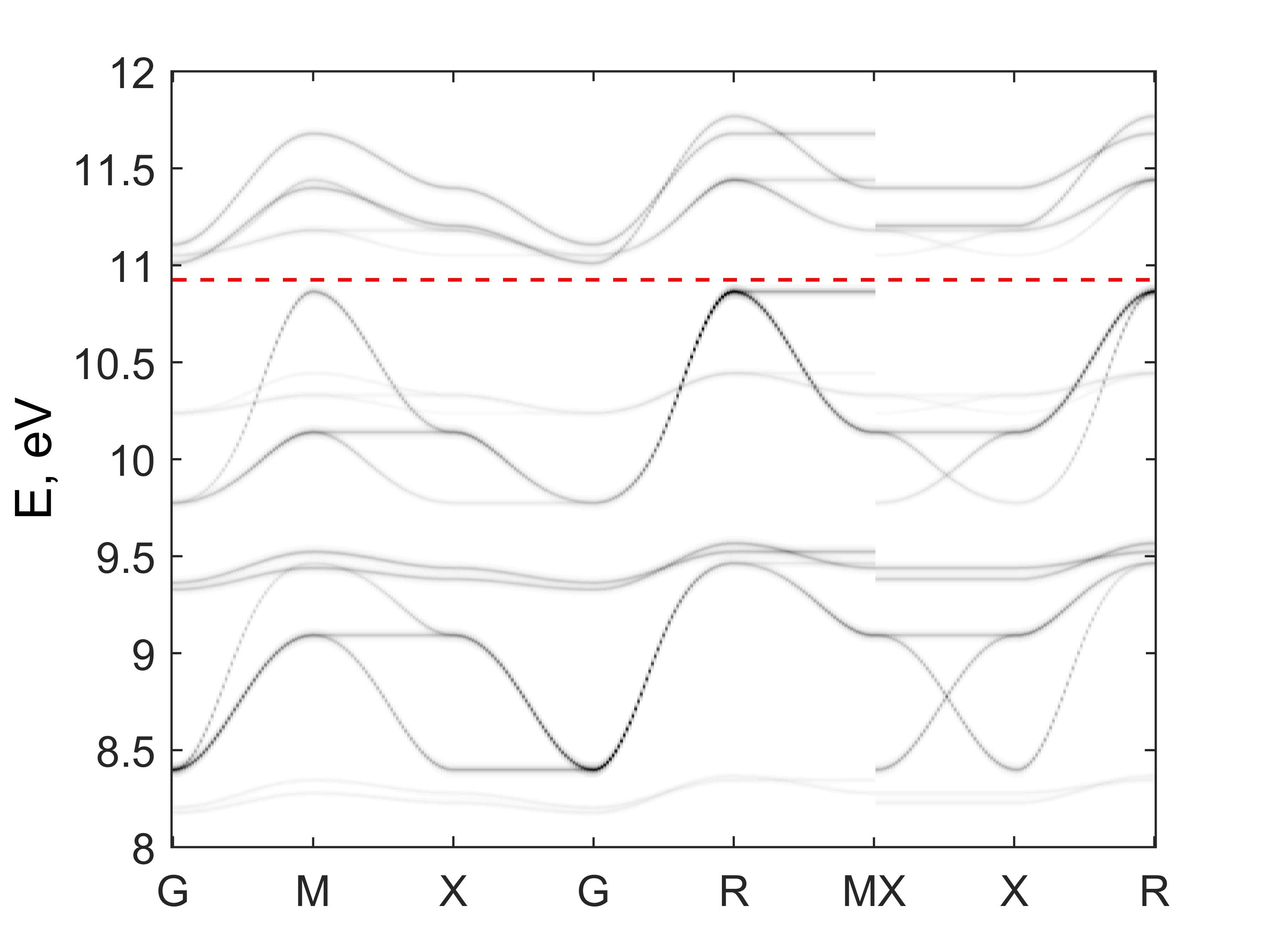} (a)}
\end{minipage}
\hfill
\begin{minipage}[h]{0.49\linewidth}
\center{\includegraphics[width=1.0\linewidth]{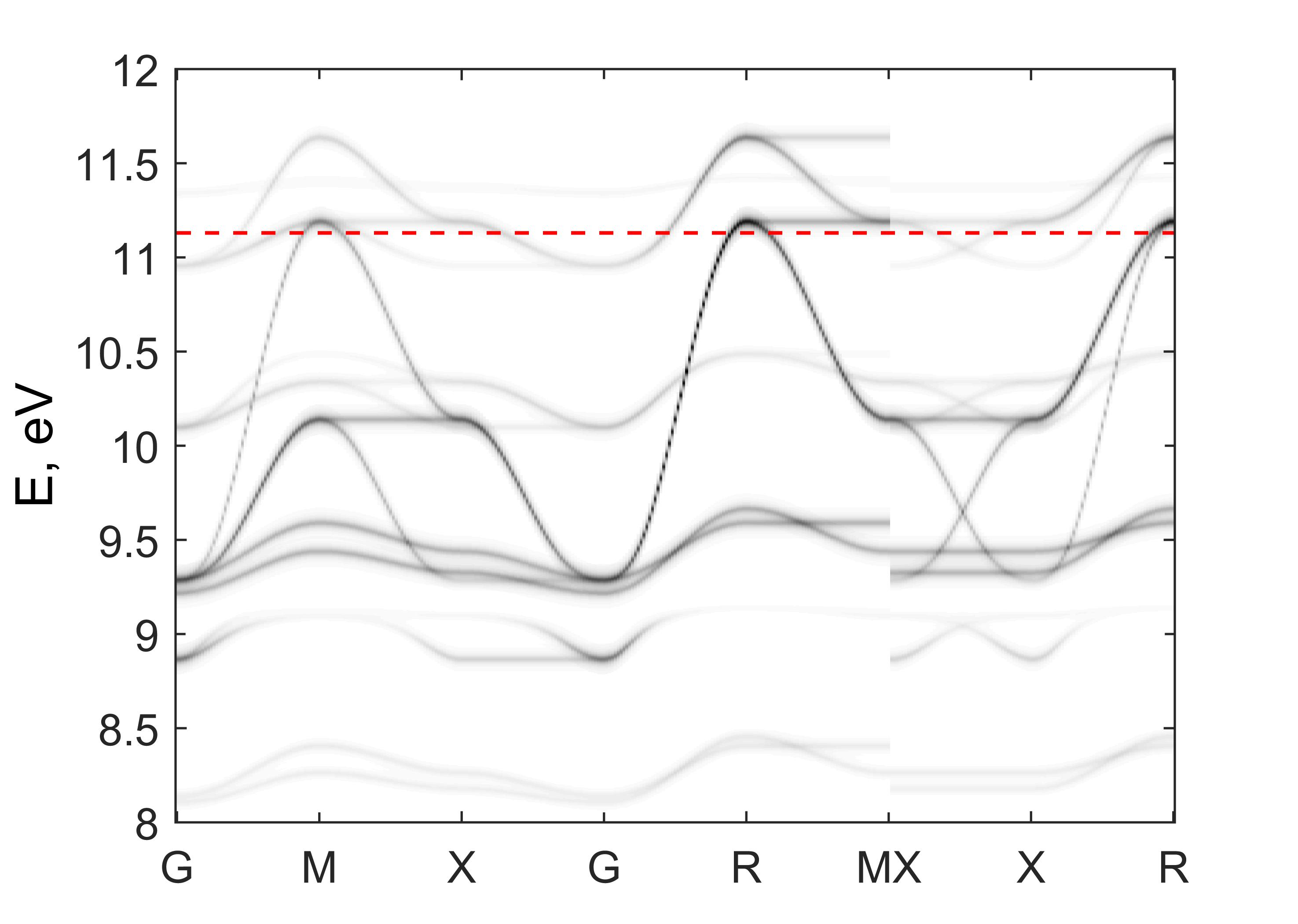} (b)}
\end{minipage}
\caption{\label{fig:3}Effect of HS optical population on LaCoO$_3$ quasiparticle spectrum. (a) in the LS- state before optical excitation, the band structure is of the insulator type; (b) quasiparticle spectrum at $t_1 = 817~t_0$ and at $t_2 = 1.71\cdot10^5$~$t_0$ (Fig.~\ref{fig:2}), the band structure is of the semimetal type at $t_1 < t < t_2$ with electrons and holes at the chemical potential. The red dashed line shows the chemical potential. $G(0,0,0)$, $M(\pi,\pi,0)$, $X(\pi,0,0)/(0,\pi,0)$, $R(\pi,\pi,\pi)$ are symmetric points of the Brillouin zone. More/less dark color of the dispersion curves corresponds to the more/less quasiparticle spectral intensity. Calculations were performed using the GTB method for the following parameter values: Racah parameters $B = 1065$~cm$^{-1}$ (0.132~eV), $C = \gamma B$, $\gamma = 4.808$; hopping parameters $t_{pd}^\sigma = 1.57$~eV, $t_{pd}^\pi = 0.84$~eV, ${t_{pp}} = 0.3$~eV; charge transfer energy ${\Delta _{CT}} = {\varepsilon _d} - {\varepsilon _p} = 2.4$~eV \cite{10}.
}
\end{figure*}

Figure~\ref{fig:3} presents the calculation results of the electronic band structure before (a) photoexcitation, in the LS- state, and after (b) at times $t_1$ and $t_2$ (Fig.~\ref{fig:2}), corresponding to an occupancy of the HS- state of 85\%. In the time interval $t_1 \le t \le t_2$, a semimetallic state (b) is realized, in contrast to the dielectric ground state (a).

To conclude, we have revealed magnetization temporal oscillations and a complex multiscale time dynamics relaxation of the magnetization, HS- state population, and the cation-anion bond length in strongly correlated LaCoO$_3$ systems. In the process of relaxation, a dielectric -- semimetal -- dielectric transition was discovered. We hope that this work will stimulate further experimental studies of the ultrafast time dynamics of magnetically ordered and non-ordered systems with spin crossover.

\begin{acknowledgments}
The authors thank the Russian Scientific Foundation for the financial support under the grant 18-12-00022.
\end{acknowledgments}


\end{document}